\documentclass[journal]{IEEEtran}
\usepackage{amsmath,amsfonts}
\usepackage{amsthm}
\usepackage{amssymb}
\usepackage{csquotes}
\usepackage{blkarray}
\usepackage{tcolorbox}
\usepackage{bm}
\usepackage{cases}
\usepackage{xcolor}
\usepackage{hyperref}
\usepackage{booktabs}
\usepackage{multirow}
\usepackage{setspace}
\usepackage{algorithm}
\usepackage[noend]{algpseudocode}

\usepackage{array}
\usepackage[caption=false,font=normalsize,labelfont=sf,textfont=sf]{subfig}
\usepackage{textcomp}
\usepackage{stfloats}
\usepackage{url}
\usepackage{verbatim}
\usepackage{comment}
\usepackage{graphicx}
\newtheorem{theorem}{Theorem}

\newtheorem{remark}{Remark}
\newtheorem{lemma}{Lemma}
\newtheorem{definition}{Definition}

\usepackage{caption}
\usepackage[noadjust]{cite}

\DeclareMathOperator*{\argmin}{arg\,min}
\DeclareMathOperator*{\arginf}{arg\,inf}
\newcommand{\btt}[1]{{\fontfamily{lmtt}\selectfont #1}}

\usepackage{float}
\floatstyle{ruled}
\newfloat{model}{thp}{lop}
\floatname{model}{Model}

\definecolor{niceblue}{rgb}{0, 0.5, 1.0}
\definecolor{niceblue}{rgb}{0.125, 0.406, 0.852}

\hypersetup{
    colorlinks=true,
    linkcolor=niceblue,
    filecolor=magenta,      
    urlcolor=niceblue,
    citecolor=niceblue}
\usepackage{balance}
\begin{document}
% \title{Tighter Global Performance Guarantees for Learned Piecewise Linear Power Flow Models}
%\title{Verifiably Robust Regions of DC-OPF}
\title{Identifying the Smallest Adversarial Load Perturbation that Renders DC-OPF Infeasible}
%\title{Global Performance Guarantees for Piecewise Linear Power Flow Models Learned from Data}
\author{Samuel Chevalier,~{\textit{Member, IEEE}}, William A. Wheeler,~{\textit{Member, IEEE}}
\thanks{This work was supported in part by the Leahy Institute for Rural Partnerships at the University of Vermont.

S.~Chevalier and W.~Wheeler are with the Department of Electrical and Biomedical Engineering, University of Vermont, Burlington, VT, USA
{\{schevali, wwheele1\}@uvm.edu}.}
%\thanks{Samuel Chevalier is supported by the HORIZON-MSCA-2021 Postdoctoral Fellowship Program, Project \#101066991 – TRUST-ML. Spyros Chatzivasileiadis is supported by the ERC Starting Grant VeriPhIED, Grant Agreement No. 949899. Both authors are with the Department of Wind and Energy Systems at the Technical University of Denmark (DTU), Kongens Lyngby, Denmark. Emails: \{schev,spchatz\}@dtu.dk.}}
}

%\markboth{Journal of \LaTeX\ Class Files,~Vol.~18, No.~9, September~2020}%

\maketitle

\begin{abstract}

What is the globally smallest load perturbation that renders DC-OPF infeasible? Reliably identifying such ``adversarial attack" perturbations has useful applications in a variety of emerging grid-related contexts, including machine learning performance verification, cybersecurity, and operational robustness of power systems dominated by stochastic renewable energy resources. In this paper, we formulate the inherently nonconvex adversarial attack problem by applying a parameterized version of Farkas' lemma to a perturbed set of DC-OPF equations. Since the resulting formulation is very hard to globally optimize, we also propose a parameterized generation control policy which, when applied to the primal DC-OPF problem, provides solvability guarantees. Together, these nonconvex problems provide guaranteed upper and lower bounds on adversarial attack size; by combining them into a single optimization problem, we can efficiently ``squeeze" these bounds towards a common global solution.
%in many tested cases, we are able to identify upper and lower bounds that are equal in value, indicating that a global solution to the problem has been found. 
We apply these methods on a range of small- to medium-sized test cases from PGLib, benchmarking our results against the best adversarial attack lower bounds provided by Gurobi 12.0's spatial Branch and Bound solver. 

\end{abstract}


\section{Introduction}

\IEEEPARstart{D}{C} Optimal Power Flow (DC-OPF) is an important tool used by transmission system operators around the world~\cite{kirschen2018fundamentals}. Using a simplified power flow assumption, DC-OPF seeks to serve load at minimal generation cost subject to generation, line flow, and power balance limits. DC-OPF constraints are furthermore embedded in a number of critical power grid operational tools, like security constrained economic dispatch (SCED)\cite{sced}, security constrained Unit Commitment (SCUC)~\cite{scuc}, and optimal transmission switching~\cite{ots, molzahn2019survey}.

Power grids operate with a non-empty feasible region. As system loads change throughout the day, the size and shape of the feasible region changes correspondingly, hopefully never collapsing to the empty set. When the feasible region is known \textit{a priori} to be empty, recent work has focused on solving \textit{infeasibility} problems which restore operational feasibility through minimized control action. Using an equivalent circuit formulation of an AC power grid,~\cite{marko_inf} found the smallest extraneous current injection that yielded a feasible power flow solution. This approach was extended to three-phase distribution grid infeasibility in~\cite{inf_dist}, and combined Transmission and Distribution grid infeasibility in~\cite{infTD}. Other recent work in this domain has used Machine Learning (ML) to diagnose and restore OPF feasibility through predictive modeling~\cite{mlInf}.

\begin{figure}
\begin{center}
\includegraphics[width=0.95\columnwidth]{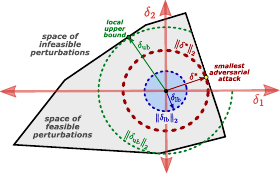}
\caption{Depicted is the globally smallest load perturbation, $\delta^*$, which renders DC-OPF infeasible along with a local solution, $\delta_{\rm ub}$, which upper bounds the attack size. Minimizing distance to infeasibility is a nonconvex problem.}
\label{fig:delta}
\end{center}
\end{figure}

This paper asks a related, but fundamentally different, question: what is the globally smallest load change which will cause the DC-OPF feasible solution space to vanish?\footnote{``Smallest'' must be defined with respect to some metric. In this paper, we consider the standard Euclidean metric. If we were to consider the different costs for perturbations, or the probabilities of different stochastic injections, a different metric could be used.} At first consideration, this would seem to be an easier problem than AC infeasibility analysis. DC-OPF is a Linear Program (LP), efficiently solvable in polynomial time. However, from the interior of the LP's feasible space, finding the smallest distance to infeasibility is a nonconvex problem. This is illustrated in Fig.~\ref{fig:delta}, where $\delta_{\rm ub}$ is a local solution to the problem (moving left or right will require larger perturbations for infeasibility), but $\delta^*$ is clearly the global solution. 

Finding the smallest load perturbation that yields operational infeasibility, assuming a binary-fixed generator commitment schedule, has a plethora of potential use cases. For example, it has applications in robust network operations, to ensure network operation isn't critically exposed to stochastic load and renewable fluctuations~\cite{robust_cr,adv_security}; in cybersecurity, to defend again stealthy MadIoT attacks~\cite{madiot}; and in ML performance verification, to ensure ML models don't push networks into regions with diminished feasibility margins~\cite{proxy_quality,rl_isone}. Inspired by other works~\cite{adv_security, adv_stochastic}, we formulate this as an ``adversarial attack" problem, where the smallest load perturbation which yields infeasible network constraints is considered an attack. Of course, renewable energy fluctuations are not literal attacks on a network, but they do challenge its reliability in an attack-like fashion. 

There is vast literature on finding OPF solutions that are robust to various network perturbations and contingencies. The specific problem we focus on in this paper, however, is \textit{broader} than the problems posed by the robust power flow literature. Our problem solution is completely invariant to the  dispatch of the power system -- we want to understand which perturbations generators can(not) feasibly respond to, rather than, ``is there a robust dispatch solution?" for a particular set of perturbations. To the authors' knowledge, this perspective of targeting the smallest perturbation to infeasibility in OPF problems is not represented in the literature. 

\subsection*{Related research works}
The following paragraphs review our paper's connections to the adjacent topics of adversarial attacks, convex restrictions, chance constraints, and robust optimization.

\textit{Adversarial attacks:} In the spirit of adversarial network attacks, \cite{adv_security} proposed an adversarial ML training framework to find AC-OPF solutions which were robust against security constraint ``attacks". Adversarial robustness was also employed in~\cite{adv_stochastic} to find AC-OPF solutions which were robust against stochastic load variations. Both of these papers considered the problem of finding robust generator dispatch solutions, in contrast to our focus of identifying perturbations that render the overall problem infeasible.

\textit{Convex restrictions:} From a more conventional optimization perspective, other works have exploited convex restrictions to find regions of guaranteed operational feasibility. For example,~\cite{Nawaf_CIA,cia_beyza} proposed the use of convex inner approximations, applied to the dist-flow equations, to identify safe loadability regions in radial distribution grids. Exploiting Brouwer's fixed-point theorem,~\cite{cr} used convex restrictions of AC power flow to determine regions of guaranteed solvability in meshed transmission networks. This work was extended, via robust convex restrictions, to robust OPF applications in~\cite{robust_cr}, where load uncertainty was modeled explicitly using chance constraints. A special class of convex restrictions, known as polyhedral restrictions, was proposed in~\cite{poly_rest}, where the feasible region of a distribution grid AC-OPF was conservatively bounded by polyhedra. More recently, \cite{dc_distanceToInf} defined a feasibility metric based on voltage stability in DC grids with constant power loads. The $p$-norm distance to infeasibility was analyzed using linear and bilinear matrix inequalities. 

Convex restrictions are not directly applicable to DC-OPF formulations, since the DC power flow model is already convex, and the solvability problem can be directly cast as a linear inequality ($\exists x:Ax\le b,\;\forall b\in {\mathcal B}$?). Absent power flow nonlinearities, convex restriction approaches try to find an operating point $x^*$ which maximizes the size of the set ${\mathcal B}$. However, these approaches tend to embed assumptions about how generators respond to load changes (see the participation factors in eq.~(6) from~\cite{robust_cr}, for example), and their primary goal is to overcome the problem of nonlinearity. Our approach, which is applied the DC-OPF problem, doesn't a priori embed assumptions about how generators respond: we find the optimal generator response policy which maximizes the size of the allowable load perturbations. In summary, while most convex restriction approaches look for the largest convex power flow solvability region in the feasible space, given some generator response policy, we look for the smallest attack that renders infeasibility\footnote{A convex restriction could, in theory, be directly applied to the nonconvex adversarial attack problem proposed in Sec.~\ref{sec:adv_attacks}. However, this would result in a conservative upper bound on the minimization problem with no guarantee that \textit{smaller} attacks cannot exist, rendering the approach futile.} . Furthermore, we do so in a manner which optimizes across variable affine generation control policies.

\textit{Chance constrained optimization}: Given operational uncertainty, chance constrained methods find a dispatch solution which respects all network constraints with a probability that is greater than or equal to some specified value~\cite{ROALD2023108725} (e.g., 95\%). For example,~\cite{bienstock2014chance} formulated a chance-constrained DC-OPF problem, where variable wind forecasts were used to parameterize generation uncertainty. A proposed affine control law dictated generation response to load imbalance. Other methods use more complex recourse control policies (i.e., control policies that map from the uncertainty to decision variables), such as polynomial and piecewise linear policies~\cite{ROALD2023108725}. Alternatively, in \cite{ccdcopf}, the authors propose a chance constrained DC-OPF problem where generation response is dictated by a model which is trained on historical Automatic Generation Control (AGC) data. A joint DC-OPF chance constrained problem, where all chance constraints are simultaneously satisfied, was solved in~\cite{DC_line} using an affine control policy. While chance constrained methods ensure an operating point is robust to specified distributional uncertainties, they do not aim to identify the smallest attack that results in infeasibility.

\textit{Robust optimization:} Finally, robust optimization (i.e., worst-case optimization) frameworks seek to ensure a solution is robust to all possible realizations of an uncertain constraint parameter~\cite{ROALD2023108725}. (e.g., load uncertainty). Various uncertainty sets (e.g., elliptical, box, budgeted, polyhedral, etc.) can be considered~\cite{ROALD2023108725}. In \cite{robust_acopf}, the authors solve a robust AC-OPF problem using semidefinite programming to ensure a network dispatch is robust to renewable energy uncertainty. In \cite{bienstock2014chance}, the authors pose a chance-constrained optimization that is robust to data (i.e., estimation) errors. Robust optimization methods consider a specific set of perturbations and find operating conditions that are feasible under all of them. While closely related, the reverse question remains unanswered: finding a perturbation where no feasible operating point exists. 

\subsection*{Paper contributions}

The problems posed by the adversarial, convex restriction, chance constrained, and robust OPF literatures are related to the problem we pose in this paper, but are fundamentally different. Each of the surveyed papers focuses on the following question: given uncertainties in load, how should we set generation dispatch and control policies to find a robust dispatch solution? In our paper, we ask the following question: given generation and network constraints, what is the smallest load perturbation that engenders infeasible network operation?

\begin{comment}
how should we set generation dispatch and control policies to find a robust dispatch solution?

Given generation and network limitations, what load perturbation instantiation 

adversarial approaches: use adversarial training methods to find robust operating solutions

convex restriction: find conservative regions of guaranteed solvability for nonlinear systems

chance constraints: robust to all constraints with a specified probability

robust optimization: robust to all constraints across all uncertainty realizations

smallest delta to Infeasibility: find numerical control policy which defends a

Add

Across all surveyed techniques, none of them seek to find the smallest perturbation 

This paper formulates and solves an operational problem 
\end{comment}

Specifically, this paper offers the following contributions:
\begin{enumerate}
    \item We pose a problem new to the study of power system optimization: identify the smallest network perturbation that yields constraint infeasibility. 
    \item We directly state the nonconvex adversarial attack problem by applying a parameterized version of Farkas' lemma to a perturbed DC-OPF problem.
    \item We establish a lower bound on the smallest attack size by formulating a numerical control policy whose optimal solution provides regions of guaranteed DC-OPF solvability.
    \item Under certain assumptions, we prove that the distance between the adversarial attack (upper bound) and the defending control policy (lower bound) ``squeeze" towards a common global solution with 0 gap.
    %\item Our approach allows us to tractably find the smallest network perturbation that yields constraint infeasibility. To our knowledge, this is the first paper that provides methodology to directly solve this problem.
    %Finally, we formulate a single optimization problem which minimizes the distance between the adversarial attack (upper bound) and the defending control policy (lower bound), thus ``squeezing" the upper and lower bounds toward a common, global solution.
    %\will{We don't actually use this - if it's going in the appendix, maybe this statement should be changed? }
\end{enumerate}

In Sec.~\ref{sec:dcopf}, we review the load-perturbed DC-OPF problem. In Sec.~\ref{sec:adv_attacks}, the adversarial attack framework is proposed, using a combination of Farkas' lemma (to parameterize infeasibility) and a numerical generation control policy (to define a solvability region). Our proposed solution procedure is outlined in Sec.~\ref{sec:solution_method}. Test results are presented in Sec.~\ref{sec:tests}, and conclusions are offered in Sec.~\ref{sec:conclusion}. In this paper, upper case ($A,B,C$) denotes matrices, and lowercase ($x,y,z$) denotes vectors.

\section{DC-OPF Modeling}\label{sec:dcopf}

We consider a power network, whose graph ${\mathcal G}({\mathcal V},{\mathcal E})$ has edge set $\mathcal{E}$, $|\mathcal{E}|=n_l$, vertex set $\mathcal{V}$, $|\mathcal{V}|=n_b$, and signed nodal incidence matrix $E\in{\mathbb R}^{n_l\times n_b}$. The diagonal matrix $Y_l={\rm diag}(b)\in{\mathbb R}^{n_l\times n_l}$ has line susceptances on its diagonals, and it relates nodal injections (generation minus demand) and phase angles $\theta\in {\mathbb R}^{n_b}$ via $p_{g}-p_{d}=E^{T}Y_l E\theta$. The canonical DC-OPF problem is given in Model~\ref{model:dcopf}, where $\delta\in {\mathbb R}^{n_b}$ represents a generalized load perturbation, and $\Phi\in{\mathbb R}^{n_l\times n_b}$ is the power transfer distribution factor (PTDF) matrix. This matrix is constructed by appending a leading zero column onto the \textit{reduced} PTDF matrix:
$\hat{\Phi}\triangleq Y_{l}\hat{E}(\hat{E}^{T}Y_{l}\hat{E})^{-1}$, where $\hat{E}$ is the reduced incidence matrix (i.e., the column associated with the slack bus has been removed). While \eqref{eq: dcopf} does not include, e.g., reserve or security constraints, the methods proposed in this paper can directly accommodate these additions. Generally, our methods accommodate any additions that can be captured by linear inequality constraints. 

\begin{model}[h]
\caption{\hspace{-0.1cm}\textbf{:} Perturbed DC-OPF (Linear Program)}
\label{model:dcopf}
\vspace{-0.25cm}
\begin{subequations}\label{eq: dcopf}
\begin{align}
\underset{\underline{p}_{g}\leq p_{g}\leq\overline{p}_{g}}{\min}\quad & c_{g}^{T}p_{g}\\
{\rm s.t.}\quad & 1^{T}p_{g}=1^{T}(p_{d}+\delta)\\
 & \underline{p}_{f}\leq\Phi(p_{g}-p_{d}-\delta)\leq\overline{p}_{f}.
\end{align}
\end{subequations}
\vspace{-0.5cm}
\end{model}

%\begin{subequations}\label{eq: dcopf}
%\begin{align}
%\underset{\underline{p}_{g}\leq p_{g}\leq\overline{p}_{g}}{\min}\quad & c_{g}^{T}p_{g}\\
%{\rm s.t.}\quad\quad & 1^{T}p_{g}=1^{T}(p_{d}+\delta)\\
% & \underline{p}_{f}\leq\Phi(p_{g}-(p_{d}+\delta))\leq\overline{p}_{f},
%\end{align}
%\end{subequations}

To characterize the feasible space of Model~\ref{model:dcopf} using inequalities, we can solve the linear power balance equation, thus eliminating a single slack generator from the decision variable set.\footnote{The DC-OPF solution, and all results in the paper, are invariant to the selection of the slack generator.} For a given load perturbation $\delta$, the feasible space associated with \eqref{eq: dcopf}, can be characterized with inequalities via
\begin{align}\label{eq: feas_space}
\mathcal{F}(\delta)=\left\{ p\,:\,Ap+B\delta+c\le0\right\}.
\end{align}
where we have defined $p\triangleq \hat{p}_g$ as the reduced generation vector to simplify notation, and where the definitions of $A$, $B$, and $c$, along with the transformation from \eqref{eq: dcopf} to \eqref{eq: feas_space}, are given in Appendix \ref{AppA}. 
%We choose to explicitly solve the power balance equality constraint, rather than convert it into two inequalities, so that we can find solutions $p$ such that \eqref{eq: feas_space} has non-zero margins to infeasibility: $Ap+B\delta+c<0$. This characteristic will be numerically exploited in later sections.

\section{DC-OPF Adversarial Attacks}\label{sec:adv_attacks}

We consider a simple question: what is the smallest load perturbation $\delta$ which can engender an infeasible set of network constraints? We denote $\delta^{*}$ as the globally greatest lower bound on perturbations that yield an empty feasible space\footnote{We define $\delta^*$ via the infimum since there is no minimum. The boundary of the feasible region is also feasible, so for any infeasible point, we can always find a new one even closer to the boundary (hence, no minimum). Formally, the vector $\delta^*$ is the limit of some sequence of vectors $\{\delta_j\}$ for which $\Vert\delta_j\Vert$ converges to $\inf_\delta \Vert\delta\Vert$.}:
\begin{subequations}\label{eq: feas_space_empty}
\begin{align}
\delta^{*}\triangleq\arginf_{\delta}\quad & \delta^{T}\delta\\
{\rm s.t.}\quad & \mathcal{F}(\delta)=\ensuremath{\emptyset}.\label{eq: empty_set}
\end{align}
\end{subequations}
Per \eqref{eq: empty_set}, if we apply load perturbation $\delta^{*}$ to a power grid, there is no value of $p$ which can satisfy $Ap+B\delta^{*}+c\le0$.

\begin{definition}[Adversarial attack]
    %We refer to $\delta^{*}$, which is the smallest load perturbation that can generate an infeasible set of network constraints, as the \textbf{adversarial attack}.
    We refer to $\delta^{*}$, which provides the greatest lower bound on load perturbations that can generate an infeasible set of network constraints, as the \textbf{adversarial attack}. 
    \footnote{For clarity, we refer to $\delta^*$ as the smallest adversarial attack, although, strictly speaking, the perturbation $\delta^*$ remains on the feasible boundary.}
\end{definition}
%\will{we say "smallest adversarial attack" in many places in the paper, and I don't want to change them all to "greatest lower  bound" because I think it makes it hard to read. so I thought we can just say here that we're going to call it the smallest attack}
To formulate the infeasibility constraint \eqref{eq: empty_set} explicitly, we exploit the fact that dual unboundedness corresponds to primal infeasibility~\cite{boyd2004convex}. Accordingly, we construct the Lagrange dual associated with the feasible space of~\eqref{eq: feas_space}:
\begin{align}
\max_{\mu\ge0}\;\min_{p} &\;  \mu^{T}\left(Ap+B\delta+c\right).
\end{align}
To engender dual unboundedness, we require: ($i$) that the term in the Lagrangian associated with the primal variable is driven to zero; and ($ii$) strict positivity of the remaining term, which grows with $\mu$ without bound.
By replacing the infeasibility condition in \eqref{eq: empty_set} with these two constraints (and $\mu \ge 0$), we have an explicit adversarial attack model, given in Model~\ref{model:minimum_perturbation}. %The parameter $\epsilon$ is chosen large enough to be numerically significant, but small enough as to not affect the solution in any practical way (e.g., $\epsilon=10^{-4}$).
\begin{model}[h]
\caption{\hspace{-0.1cm}\textbf{:} Smallest Adversarial Attack to DC-OPF Infeasibility}
\label{model:minimum_perturbation}
\vspace{-0.25cm}
\begin{subequations}\label{eq: farkas}
\begin{align}
\delta^{*}\triangleq\argmin_{\mu\ge 0, \delta}\quad & \delta^{T}\delta\\
{\rm s.t.}\quad & A^T \mu =0\label{eq: f1}\\
 & \mu^{T}\left(B\delta+c\right)>0\label{eq: f2}%=\epsilon,\quad 1\gg\epsilon>0\label{eq: f2}
\end{align}
\end{subequations}
\vspace{-0.5cm}
\end{model}

%\will{In this model, you've side-stepped the issue of the minimum existing by setting the buffer $\epsilon$ - which mathematically means that it's not actually the smallest perturbation to infeasibility. It seems like $\epsilon$ is introduced here for numerical/computational reasons rather than math ones. Is it better to separate math from numerics, or is it all in the context of a numerical implementation?}

Model~\ref{model:minimum_perturbation} can be interpreted as a problem which finds the minimum load perturbation that engenders network infeasibility (see Fig.~\ref{fig:delta}). This model is nonconvex due to the bilinear interaction between $\mu$ and $\delta$. While no \textit{a priori} upper-bounds on $\mu$ are inferrable, this problem has an additional degree of freedom, which lets us scale $\mu$ arbitrarily. For example, we can add the normalization constraint $1=1^T\mu$ without changing the optimal solution $\delta^*$. Despite this fact, convex relaxations of formulation \eqref{eq: farkas} tend to be very weak\footnote{In computational tests, SDP relaxations of \eqref{eq: farkas}, combined with McCormick and RLT cuts, always produced a less-than-useful lower bound of 0. Based on these results, we did not explore relaxation-based methods in this paper.}.

\begin{remark}
The primal infeasibility constraints captured by \eqref{eq: f1}-\eqref{eq: f2} are a parameterized version of \textbf{Farkas' lemma}~\cite{boyd2004convex}.
\end{remark}
Appendix \ref{AppC} reviews Farkas' lemma. Assuming the unperturbed DC-OPF problem \eqref{eq: dcopf} is solvable, $\delta=0$ cannot be a solution to \eqref{eq: farkas}. However, a feasible solution will always exist.

\begin{remark}
    Model~\ref{model:minimum_perturbation} always has a feasible solution.
\begin{proof}
    By Farkas' lemma, the primal constraint in \eqref{eq: feas_space}, and the dual constraints in \eqref{eq: f1}-\eqref{eq: f2}, are \textit{alternative} systems: exactly one system of constraints is satisfiable. There always exists a load perturbation which yields DC-OPF infeasibility. For example, set $\delta$ to violate generator dispatch limits via $1^T\delta > 1^T(\overline{p}_g-p_d)$: in this case, Model~\ref{model:minimum_perturbation} must always be feasible.
\end{proof}    
\end{remark}

Since Model~\ref{model:minimum_perturbation} is a nonconvex quadratically constrained quadratic program (QCQP), solving it to global optimality requires extensive branching and bounding and is generally very slow. However, \eqref{eq: farkas} can be locally solved with relative ease via, e.g., an interior point solver. The resulting ``incumbent" solution ${\delta}_{\rm ub}$ provides an upper bound to the global solution, while ${\delta}_{\rm lb}$ (the focus of the next subsection) is a lower bound: %\will{$\delta_{\rm lb}$ is not really defined yet?}
\begin{align}
\bigl\Vert {\delta}_{\rm lb} \bigr\Vert\le  \bigl\Vert {\delta}^* \bigr\Vert  \le \bigl\Vert {\delta}_{\rm ub} \bigr\Vert.
\end{align}
In testing, Branch and Bound (BaB) tends to find good upper bounds fairly quickly. Proving a lower bound $\delta_{\rm lb}$ for Model~\ref{model:minimum_perturbation}, however, is very hard, even on small power systems. For example, we applied Model~\ref{model:minimum_perturbation} to a 14-bus power system. After one hour of branching and bounding, Gurobi v12 still had a lower bound of 0 and an incumbent of 0.178 (which is the global solution).

The core computational challenge of this paper, therefore, hinges on providing good lower bounds for Model~\ref{model:minimum_perturbation}, i.e., guaranteeing that there can be no smaller adversarial attack than ${\delta}_{\rm lb}$. Unfortunately, convex restrictions are not directly applicable for lower bounding the solution space of \eqref{eq: farkas}.

\begin{remark}
    Since Model~\ref{model:minimum_perturbation} is a minimization, convex restrictions~\cite{cr} of the formulation result in a conservative upper bound with no guarantees on minimum adversarial attack size.
\end{remark}

Next, we introduce a primal method for bounding regions of guaranteed solvability, thus providing ${\delta}_{\rm lb}$ candidates.

\subsection{Lower bounding the adversarial attack}

To find a lower bound candidate ${\delta}_{\rm lb}$ such that $\Vert{\delta}_{\rm lb} \Vert \le \Vert {\delta}^* \Vert$, we consider the original primal constraint \eqref{eq: feas_space}. For $\delta=0$, we may choose a constant $p_0$ such that no feasibility constraint is tight: $Ap_0+c<0$. In this case, every constraint has a nonzero robustness margin. If a norm-bounded perturbation cannot violate any single constraint, then the primal system must be feasible for all perturbations within this ball. To ensure this, we consider the $i^{\rm th}$ constraint and find the \textit{minimal} perturbation that touches the constraint margin boundary by solving
\begin{subequations}\label{eq: quad_proj}
\begin{align}
{\tilde \delta}^{(i)}_{\rm lb}\triangleq\argmin_{\delta}\quad & \delta^{T}\delta\\
{\rm s.t.}\quad & a_{i}^{T}p_0+b_{i}^{T}\delta+c_{i}=0,
\end{align}
\end{subequations}
where $a_{i}^T$ is the $i^{\rm th}$ row of $A$ and $b_i^T$ is the $i^{\rm th}$ row of $B$.
The solution to \eqref{eq: quad_proj}, which represents an optimal projection of $\delta=0$ onto an equality constraint, is given by \eqref{eq: pi_solution} in Appendix \ref{AppB}. If we take the smallest perturbation across all minimal constraint perturbations, we effectively \textit{lower bound} the smallest load perturbation which leads to network infeasibility:
\begin{align}\label{eq: lb_og_bound}
{\delta}_{\rm lb}=\argmin_{\tilde{\delta}_{\rm lb}^{(i)},\; i\in\mathcal{C}}\quad\bigl\Vert {{\tilde \delta}}^{(i)}_{\rm lb} \bigr\Vert _{2},
\end{align}
% \will{this equation defines $\delta_{\rm lb}$ as a scalar, but in eq 6 is is treated as a vector. Also, it looks like this symbol is not used any more in the paper. Should it be a scalar or a vector?}
where $\mathcal C$ is the constraint set. We call this a \textit{lower bound} on the adversarial attack, rather than a global solution, because the decision variables $p$ have remained fixed at $p_0$. In other words, we have implicitly assumed that all load perturbation is picked up by the slack bus. When a load perturbation is applied to a real network, these decision variables are free to move according to operator directions. In the next subsection, we envision a parameterized control policy which allows the decision variables to move in direct response to a load perturbation, thus improving this lower bound.

\subsection{Improving the adversarial attack lower bound via parameterized control policy}

In this subsection, we improve the lower bound provided in \eqref{eq: lb_og_bound} by adding in an explicit generation response control policy (i.e., allowing non-slack generation to vary in response to load perturbations). Generation response control policies have been proposed in many papers, e.g., in Bienstock~\cite{bienstock2014chance}. We note that our control policy is not needed, nor intended, to be used to control generation response in the actual system; 
%\will{why not? isn't this the one that guarantees the largest safe perturbation? If an arbitrary different one were implemented, wouldn't the robustness be lost? -- also I'm not really sure why the rank-1 examples are helpful. Can we not just jump directly to saying we are considering a general linear control policy $p_0+G\delta$?} 
it is simply a numerical crutch which allows us to find successively larger regions of guaranteed solvability. In other words, it allows us to prove that a feasible solution exists which the operators could find using their conventional DC-OPF tools. 

We may envision various types of explicit control policies. For example:
\begin{itemize}
    \item A distributed slack policy may equally distribute load changes to all generators according to $1^{T}(\delta-\delta_0)/n_{g}$;
    \item Proportional distributed slack may distribute load changes to generators according to $p_i1^{T}(\delta-\delta_0)/\sum p_i$.
\end{itemize}
We refer to both of these policies as ``rank-1" control policies, in the sense that they can be captured by a rank-1 update to the PTDF matrix, as shown by an example in Appendix \ref{AppD}.
%\textbf{Example 1:} \textit{Rank-1 Control Policy. }Assuming a uniform distributed slack model, injection perturbations ($p-p_0$) are mapped to nodal generator responses via 
%\begin{align}
%\Phi\left(p+1\frac{1^{T}(p-p_{0})}{n}\right)=\underbrace{\Phi\left(I+\frac{11^{T}}{n}\right)}_{\text{rank-1 PTDF update}}p-\underbrace{\Phi\frac{11^{T}}{n}p_{0}}_{\rm bias}.
%\end{align}
%Thus, distributed slack implicitly applies a rank-1 update to the PTDF matrix. \hfill $\qed$
While simple to construct, rank-1 control schemes are severely limited in the range of control policies they can encode. 

In this paper, we propose an arbitrary, rank $n_g-1$ control policy $G\in \mathbb{R}^{(n_g-1)\times n_b}$, which maps load perturbations $\delta$ to generation responses. We construct this policy \textit{around} a feasible operating point $p_0$, leading to the feasibility problem
\begin{align}
Ap_0+\left(AG+B\right)\delta+c\le0,
\end{align}
where $p_0$ is the base operating point, and $G\delta$ is the perturbation of the generation; implicitly, generation is updated according to the control law 
\begin{align}
p=p_0+G\delta, 
\end{align}
which we refer to as an affine control policy.
Using this embedded control law, we can re-solve for the perturbation $\delta_{\rm lb}^{(i)}$ that lower bounds infeasibility of the $i^{\rm th}$ system constraint:
%minimal load perturbation, associated with a given system constraint $i$, which yields network infeasibility: 
\begin{subequations}\label{eq: p_control}
\begin{align}
\delta^{(i)}_{\rm lb}\triangleq\argmin_{\delta}\quad & \delta^{T}\delta\\
{\rm s.t.}\quad & a_{i}^{T}p_0+a_{i}^{T}G\delta+b_{i}^T\delta+c_{i}=0.
\end{align}
\end{subequations}
The solution to \eqref{eq: p_control}, derived in Appendix \ref{AppB}, is given by \eqref{eq: param_pert_size}. The smallest solution norm across all constraints will lower bound the adversarial attack size. Our goal, therefore, is to find a control policy which maximizes the smallest perturbation:
\begin{align}
\max_{G,p_0}\;\min_{i\in\mathcal{C}}\quad\bigl\Vert \delta_{{\rm lb}}^{(i)}\bigr\Vert _{2}^{2}.
\end{align}
In Model~\ref{model:epi}, we rewrite this using the epigraph trick and the explicit solution for the perturbation norm from \eqref{eq: param_pert_size}.

\begin{model}[h]
\caption{\hspace{-0.1cm}\textbf{:} Control Defense with Largest Feasibility Guarantee}
\label{model:epi}
\vspace{-0.25cm}
\begin{subequations}\label{eq: epi}
\begin{align}
t^* \triangleq \max_{t,G,p_0}\quad & t\\
{\rm s.t.}\quad & t\le\frac{\left(a_{i}^{T}p_0+c_{i}\right)^{2}}{\left(G^{T}a_{i}\!+\!b_{i}\right)^{T}\left(G^{T}a_{i}\!+\!b_{i}\right)},\;\forall i\in\mathcal{C}\label{eq: tle}\\
&Ap_0 + c\le 0.\label{eq: local_const}
\end{align}
\end{subequations}
\vspace{-0.5cm}
\end{model}
Constraint \eqref{eq: local_const} is necessary; without it, the optimizer has no motivation to choose a base operating point $p_0$ which is feasible, meaning it may find a control policy which indeed enables large margins to the feasibility boundary, but in some cases, from the wrong side of the inequality.

\begin{lemma}
    There is no adversarial attack smaller than $t^*$.
\begin{proof}
    Injection vector $p_0$ is feasible by \eqref{eq: local_const}, so there is a nonnegative margin between each constraint and the feasibility boundary when $\delta=0$. For each constraint, the globally smallest perturbation $\delta$ which leads to 0 feasibility margin can be found by solving \eqref{eq: ld_gp}, which is convex. $t^*$ is the norm of the smallest of all of these perturbations, so no adversarial attack smaller than $t^*$ can yield network infeasibility.
\end{proof}
\end{lemma}

Notably, \eqref{eq: epi} is a hard, nonconvex program which generally requires the use of branching and bounding to find a globally optimal solution. However, the value of any non-optimal, or local, solution can still provide a helpful lower bound to Model~\ref{model:minimum_perturbation}. To state this explicitly, for a given control policy and nominal operating point satisfying $Ap_0+c\le 0$, we define $\tilde{t}\le t^*$ as
\begin{align}\label{eq: t_tilde}
\tilde{t}\left(p_0,G\right)=\min_{i\in\mathcal{C}}\left\{ \frac{\left(a_{i}^{T}p_0+c_{i}\right)^{2}}{\left(G^{T}a_{i}+b_{i}\right)^{T}\left(G^{T}a_{i}+b_{i}\right)}\right\}.
\end{align}
This lower bounds the adversarial attack size via %\will{$\tilde{t}$ is feasible right? so this inequality should be strict}
\begin{align}\label{eq: cut_rsoc}
\tilde{t}\left(p_0,G\right) < \left\Vert \delta^{*}\right\Vert _{2}^2.
\end{align}
%This cut provides a critically useful lower bound for the Branch and Bound Solver. 
Next, 
%we combine the search for $\delta^*$ and $t^*$ into a single optimization problem. First, however, 
we consider the relationship between $t^*$ and $\left\Vert \delta^{*}\right\Vert^2 _{2}$.

\subsection{Optimal parameterized control policies can tightly bound the adversarial attack region}
In this subsection, we ask the question: can the solvability region carved out by the parameterized control policy ($G$, $p_0$) be as large as the size of the adversarial attack $\delta^{*}$? We first note that $\left\Vert \delta^{*}\right\Vert^2_{2}$, which is the smallest adversarial attack, trivially \textit{upper bounds} the largest control defense radius $t^{*}$, where there is guaranteed feasibility.
% \will{This is saying that the smallest attack size (causing infeasibility) upper bounds the largest control defense radius (guaranteed feasibility). I think we should add language like this -- probably is obvious to you but I think the triviality is clearer thinking about the concepts, whereas I have to remind myself what the variables mean}
\begin{remark}
    $t^{*}$ cannot be larger than $\left\Vert \delta^{*}\right\Vert^2_{2}$. If it was, this would imply the existence of a control policy which can find feasible DC-OPF solutions for infeasible  perturbations $\delta^{*}$. %larger than $\left\Vert \delta^{*}\right\Vert^2 _{2}$ (but $\le t^{*}$). By Farkas' lemma, as applied in Model~\ref{model:minimum_perturbation}, such solutions cannot exist, since $\delta^{*}$ is an infeasible perturbation.
\end{remark}

To make an even stronger claim, we show an equivalence between $t^*$ and the smallest adversarial attack to infeasibility $\Vert\delta^*\Vert_2^2$ under certain conditions. 
We define $\Delta$  as the set of perturbations bounded by the adversarial attack $\delta^{*}$ as an unreachable supremum:
\begin{align}\label{eq: Delta}
\Delta\triangleq\left\{ \delta:\;\left\Vert \delta\right\Vert _{2}<\left\Vert \delta^{*}\right\Vert _{2}\right\}.
\end{align}
By this definition, there exists a feasible solution for all perturbations in $\Delta$:
\begin{align}\label{eq: primal_solvability}
\forall\delta\in\Delta,\exists p:Ap+B\delta+c\le0.
\end{align}
We make the following conjecture:
\begin{align}\label{eq: low_rank_solvability}
\exists p_{0},G:Ap_{0}+\left(AG+B\right)\delta+c\le0,\forall\delta\in\Delta,
\end{align}
i.e., there exists a control policy whose feasible space includes all of $\Delta$.
%In other words, there must exist a control policy which has an equivalently large feasible space as $\Delta$ in \eqref{eq: primal_solvability}. 
%Proving this conjecture in the general case is challenging, if possible. Instead, 
While we are uncertain if this conjecture holds in general, 
we provide a proof for a slightly restricted case, where we assume feasibility of the extreme points associated with a bounding simplex $\mathcal S$. 

\begin{definition}[Simplex]
A \textbf{simplex} $\mathcal S$ in $n$-dimensional space is the convex hull of $n+1$ vertices $\delta_i \in {\mathbb R}^n$:
\begin{align}\label{eq: simplex}
\mathcal{S}=\left\{ \sum_{i=1}^{n+1}w_{i}\delta_{i} \;|\; w_{i} \ge0\,\forall i, \; {\rm and} \; \sum_{i=1}^{n+1}w_{i}=1\right\}. 
\end{align} 
\end{definition}

Given any two simplexes of the same dimension, there exists an invertible affine transformation between them~\cite{tymchyshyn2019beginner}. Equivalently, given $s\in \mathcal{S}$, the convex combination of $\delta_i$ yielding $s = \sum_i w_i \delta_i$ is unique.\footnote{The unique weights are the output of an affine map onto the simplex whose $n+1$ vertices are the $n$ unit vectors plus the origin. It is bounded by $\sum_{i=1}^n w_i \le 1$, all $w_i$ are in $[0, 1]$, and the remaining weight $1-\sum_{i=1}^n w_i$ is assigned to the vertex at the origin.} %To eventually show a restricted equivalence between $t^*$ and the adversarial attack bound, 
We offer the following minor extension to this classical result.

\begin{lemma}\label{lem: simp}
    There exists a unique affine transformation $[G\; p_0]$ from a simplex of $n$-dimensions to the convex hull of $n+1$ vertices in $m$-dimensional space.
\begin{proof}
    Collect the simplex vertices $\delta_i$ into $D$ and the potentially non-simplex (unless $m=n$) vertices $p_i$ into $P$:
    \begin{align}
    D & =\left[\begin{array}{cccc}
    \delta_{1} & \delta_{2} & \cdots & \delta_{n+1}\end{array}\right],\;\delta_{i}\in\mathbb{R}^{n}\\
    P & =\left[\begin{array}{cccc}
    p_{1} & p_{2} & \cdots & p_{n+1}\end{array}\right],\;p_{i}\in\mathbb{R}^{m}.
    \end{align}
    We posit the affine map $P=GD+p_0\mathbf{1}^T$, where $G\in {\mathbb R}^{m\times n}$ and $p_0 \in {\mathbb R}^{m}$ are unknown. We 
    introduce the $(n+1)$-dimensional vectors $\hat{\delta}_i = [\delta_i \,; 1] $  to express the map as
    %We now append these unknowns, and we append ones onto the bottom row of $X$ to get
    \begin{align}\label{eq: dhat}
P=\left[G\;p_0\right]\underbrace{\left[\begin{array}{cccc}
    \delta_{1} & \delta_{2} & \cdots & \delta_{n+1}\\
    1 & 1 & 1 & 1
    \end{array}\right]}_{\hat{D}}.
    \end{align}
    Assuming no $\delta_i$ is a convex combination of the others (i.e., no co-linearity of three points, etc.), the matrix $\hat{D}$ is invertible. Thus, the desired affine transformation is given by 
    \begin{align}\label{eq: M}
        [G\; p_0] = P\hat{D}^{-1}.
    \end{align} This result holds generally for $m\ne n$. Per the definition of the simplex \eqref{eq: simplex}, a new point $\delta_{\rm new}$ on the interior of the simplex may be expressed as 
    \begin{align}
    \delta_{\rm new}=\sum_{i=1}^{n+1}w_{i}\delta_{i},\; w_i\ge 0, \sum w_i=1.
    \end{align}
    This relationship is represented in our augmented space ($\hat{\delta}_{{\rm new}}=[\delta_{{\rm new}}\,; 1]$) by the matrix equation
    \begin{align}
        \hat{\delta}_{\rm new} = \hat{D} w,\; w_i\ge 0.
    \end{align}
    %yielding weights $w = \hat{D}^{-1} \hat{\delta}_{\rm new}$.
    Applying the transformation $[G\; p_0]$ to ${\hat \delta}_{\rm new}$ yields
    \begin{subequations}\label{eq: ch_proof}
    \begin{align}
    %G\hat{\delta}_{\rm new} & =G\sum_{i=1}^{n+1}w_{i}\hat{\delta}_{i}\\
    %\hat{y}_{\rm new} & =\sum_{i=1}^{n+1}w_{i}(G\hat{\delta}_{i})=\sum_{i=1}^{n+1}w_{i}\hat{y}_{i}.
    p_{\rm new}=[G\; p_0]\hat{\delta}_{\rm new} & =P\hat{D}^{-1} \hat{D} w \\
                       %& =Pw \\
                       & =\sum_{i=1}^{n+1} {p}_i w_i. 
    \end{align}
    \end{subequations}
    %\will{three lines might be overkill in eq 25}
    %\will{this totally works, but it's somehow unsatisfying to me to do $\sum_i w_i Y X^{-1} \hat{x}_i$ when we could just do $YX^{-1} \hat{x}_{new}$. I would opt to establish that the convex weights are uniquely determined by $w=\hat X^{-1}\hat x$ and then construct $y_{new}$ from the weights.}
    Thus, the derived affine transformation maps any point on the interior of the simplex into the convex hull of the associated $m$-dimensional points $p_i$.
\end{proof}
\end{lemma}

Working towards the conjecture in \eqref{eq: low_rank_solvability}, we consider a simplex $\mathcal{S} \subset \mathbb{R}^n$ in the space of perturbations.
If $\mathcal{S}$ contains the perturbation set $\Delta$ from \eqref{eq: Delta}, we can show that the inequality in \eqref{eq: cut_rsoc} provides a \emph{tight} lower bound to the solution of Model~\ref{model:minimum_perturbation}, $t^{*}=\left\Vert \delta^{*}\right\Vert_{2}^{2}$.
Finding conditions for the tight lower bound motivates the following theorem, which is a central result of this paper: 

\begin{theorem}\label{eq: th_simp}
    Let $\mathcal{S}$ be a simplex in $\mathbb{R}^n$. 
    There exists a linear control policy $G, p_0$ such that for all $\delta \in \mathcal{S}$, 
    \begin{equation}\label{eq:theorem_control_policy}
        Ap_0 + (AG+B)\delta + c \le 0
    \end{equation}
    if and only if 
    the extreme points of $\mathcal{S}$ have feasible solutions. 
    % \begin{equation}\label{eq:theorem_extreme_points}
    % \forall \delta_i \in  \operatorname{extreme}(\mathcal{S}),\, \exists p: Ap + B\delta_i + c \le 0.
    % \end{equation}

\begin{proof}
    %We prove both directions of this implication.
    Proving the forward direction is trivial.
    The extreme points of $\mathcal{S}$ are in $\mathcal{S}$, so \eqref{eq:theorem_control_policy} provides feasible solutions.
    % Let $p_0, G$ be a control policy given by \eqref{eq: low_rank_solvability}. Then $p = p_0 + G\delta$ is a feasible solution $\forall \delta \in \mathcal{S}$:
    % \begin{align*} 
    % A{p}+B\delta+c
    % &=
    % A{\left(p_{0}+G\delta\right)}+B\delta+c
    % \\&=
    % Ap_{0}+\left(AG+B\right)\delta+c
    % \\&\le 0.
    % \end{align*}
    
    We now prove the reverse direction by invoking Lemma~\ref{lem: simp}: 
    an affine map exists from the simplex $\mathcal{S}$ to the convex hull of $p_i \in \mathbb{R}^m$, i.e., the $n+1$ feasible solutions associated with the extreme points $\delta_i$ of $\mathcal{S}$. 
    Since our inequality constraints $Ap+B\delta+c\le0$ are linear, any point in the convex hull of a set of feasible extreme points must also be feasible. Thus, by setting the control policy ($G$, $p_0$) to the derived affine transformation \eqref{eq: ch_proof}, we have shown that a control policy exists which maps every $\delta \in \mathcal{S}$ to a feasible solution $p$.
\end{proof}
\end{theorem}

\begin{figure}
\includegraphics[width=\columnwidth]{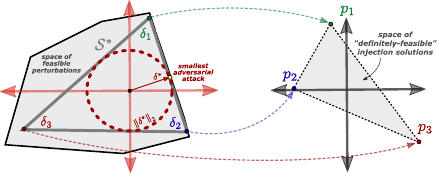}
\caption{The left plot shows the simplex, ${\mathcal S}^*$, in two dimensions, where the largest perturbation ball is inscribed inside. The extreme points of ${\mathcal S}^*$ map to feasible solutions $p$ of the injection space, which might be a space of higher or lower dimensions.}
\label{fig:delta_mapping}
\end{figure}

\begin{figure}
\includegraphics[width=\columnwidth]{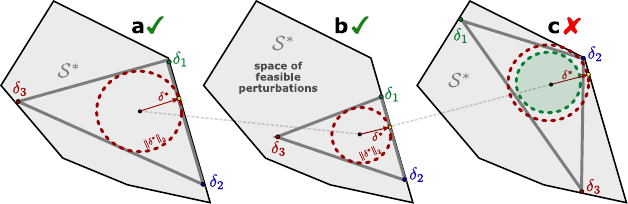}
\caption{Three different base load operating conditions (black dots). In each panel, the feasible perturbation region ${\mathcal S}^*$ is the same, but the smallest perturbation to infeasibility $\delta^*$ is different, since it depends on the base operating point. The extreme points $\delta_1$, $\delta_2$, and $\delta_3$ are selected differently in each case, so that they always fall within the space of feasible perturbations. In the first two examples, simplexes exist which fully contain the ball $\Delta$ from \eqref{eq: Delta}. However, in the third case, no simplex can be drawn which fully contains $\Delta$.}
\label{fig:delta_simplex_examples}
\end{figure}

In particular, if there exists a feasible simplex containing the perturbation set $\Delta$ from \eqref{eq: Delta}, then there exists a control policy guaranteeing feasible solutions for all $\delta \in \Delta$.
An example of such a (non-unique) simplex is illustrated in Fig.~\ref{fig:delta_mapping}.
While this condition is somewhat restrictive, we note that the theorem does not prescribe the shape of the simplex. 
As depicted in Fig.~\ref{fig:delta_simplex_examples}, simplex extreme points can be selected in clever ways which avoid infeasible perturbations. 
Of course, there exist perturbation environments $\Delta$ that are not contained by any simplex of feasible perturbations. 
In these cases, $t^*$ might become a strict lower bound on the adversarial attack size. An example of this is shown in panel \textbf{c} of Fig.~\ref{fig:delta_simplex_examples}: in this situation, no simplex with feasible extreme points can be drawn to fully contain the red ball $\Delta$.

Theorem~\ref{eq: th_simp} says nothing about how the control policy applies to points outside the simplex. Therefore, it remains \textit{possible} that $t^*$ is a tight lower bound even when the simplex does not entirely contain $\Delta$, just without guarantee. If a simplex $\mathcal{S}$ containing a ball slightly smaller than $\Delta$ exists, Theorem~\ref{eq: th_simp} guarantees that the gap between the defense radius and the attack size will be small, though not necessarily zero.

Our use of a bounding simplex, rather than, e.g., a hypercube, to prove Theorem~\ref{eq: th_simp} hinges on a special property of simplexes (which doesn't exist for general polytopes):
for every new dimension, exactly one new vertex is added to the simplex. Together, this adds one new row and one new column to matrix $\hat D$ in \eqref{eq: dhat}, keeping it both square and invertible.

%%%%%%%%%%%%%%%%%%%%%%%%%%%%%%%%%
If there is a feasible simplex $\mathcal{S}$ containing $\Delta$, Theorem~\ref{eq: th_simp} guarantees the existence of a control policy $p_0, G$ that satisfies \eqref{eq: low_rank_solvability}. Actually constructing the associated affine transformation implies knowledge of $\delta^*$, which is, of course, unknown \textit{a priori}. This control policy can be determined numerically, however, via \eqref{eq: epi}, which is a nonconvex optimization problem. In simple cases, e.g., where the perturbation space is 1-dimensional, we can analytically construct this parameterized control policy. We demonstrate this in the following example. 
%a parameterized control policy whose feasible region is the same size as the adversarial attack region.\will{what is adversarial attack region?}

\textbf{Example 2:} \textit{Control policy for scalar perturbation.} In this simplified example, which uses a generically sized linear program, we assume the applied perturbation is a scalar, $\delta\in{\mathbb R}^1$ (e.g., uniform load scaling), and $\delta^*=1$. We can now build an explicit control policy for this system. Given perturbations at the boundary of the set $\Delta$ as $\delta=\pm1$, we identify associated feasible solutions $p^{+}$ and $p^{-}$:
\begin{align*}
Ap^{+}+B\left.\delta\right|_{+1}+c & \le0\\
Ap^{-}+B\left.\delta\right|_{-1}+c & \le0.
\end{align*}
Next, we define $p^{0}=\tfrac{1}{2}(p^{+}+p^{-})$ as the nominal operating point, and $G=\tfrac{1}{2}(p^{+}-p^{-})$ as the control policy (these are computed analogously to the solution procedure given in \eqref{eq: M}). When this parameterized control policy is applied, the associated primal solution is implicitly given by 
\begin{subequations}
\begin{align}
p & =p_{0}+G\delta\\
 & =\tfrac{1}{2}(p^{+}+p^{-})+\tfrac{1}{2}(p^{+}-p^{-})\delta,\quad \delta\in[-1,1]\label{eq: conv_comb}
\end{align}
\end{subequations}
Eq.~\eqref{eq: conv_comb} represents a convex combination of primal points between $p^{-}$ and $p^{+}$. Since any convex combination of feasible points of a linear program is also feasible,~\eqref{eq: conv_comb} represents a valid control policy which satisfies \eqref{eq: primal_solvability}.\hfill $\qed$

Control policies for systems with high dimensional perturbations must generally be found numerically, which we demonstrate in the test results section.

\section{Adversarial Attack Solution Procedure}\label{sec:solution_method}
We now propose a coherent solution procedure for tightly bounding the smallest adversarial perturbation. To do this, we make three observations:
\begin{itemize}
    % Why reword these bullet points? Starting with "via incumbent solutions" makes it hard to quickly identify what each bullet point is about, and therefore what the solution concept is. Starting with "upper bound" and "lower bound" (hopefully) more quickly distinguishes the bullet point concepts.
    \item A good upper bound on attack size is provided by incumbent solutions to Model~\ref{model:minimum_perturbation} (smallest attack to infeasibility). The corresponding lower bound is loose and hard to prove.
    \item A good lower bound on attack size is provided by incumbent solutions to Model~\ref{model:epi} (control defense guarantee). The corresponding upper bound is loose.
    \item If the upper bound incumbent $\Vert\delta_{\rm inc}\Vert_2^2$ from Model~\ref{model:minimum_perturbation} and the lower bound incumbent $t_{\rm inc}$ from Model~\ref{model:epi} match, then we have a guaranteed global solution to the problem.
\end{itemize}

\begin{figure}
\begin{center}
\includegraphics[width=1.0\columnwidth]{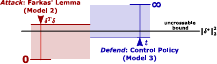}
\end{center}
\caption{Model~\ref{model:minimum_perturbation}, which seeks the smallest adversarial attack, generally struggles to raise its lower bound. Model~\ref{model:epi}, which finds a defensive control policy, has a solution $t^*$ which is potentially below $\left\Vert \delta^*\right\Vert _{2}^{2}$. Both models, however, are guaranteed to converge at the uncrossable line if there is a simplex containing the perturbation ball $\Delta$. This is exploited in Model~\ref{model:mod4} (Appendix~\ref{AppE}), which squeezes the first two model bounds together.}
\label{fig:Model_bounds}
\end{figure}

This relationship is depicted in Fig.~\ref{fig:Model_bounds}, where Model~\ref{model:minimum_perturbation} is cast as the ``attacking" upper bound, while Model~\ref{model:epi} is the ``defending" lower bound, since it is looking for a control policy which can, implicitly, defend across all bounded perturbations. If the Model~\ref{model:minimum_perturbation} upper bound and the Model~\ref{model:epi} lower bound converge to the same value, then the smallest adversarial attack has been found, even if neither model alone is able to prove its respective lower and upper bounds. 

To exploit the complementary relationship between Models~\ref{model:minimum_perturbation} and \ref{model:epi}, we present a combined model, Model~\ref{model:mod4} (given in Appendix~\ref{AppE}), whose objective is to minimize the distance between $\left\Vert \delta\right\Vert _{2}^{2}$ and $t$, pulling both constituent models simultaneously towards the uncrossable line in Fig.~\ref{fig:Model_bounds} ($t$ is pulled up, and $\delta^T\delta$ is pushed down). 
While Model~\ref{model:mod4} is an explicit statement of our strategy, we do not solve it explicitly in the test results section. Instead, we solve Models~\ref{model:minimum_perturbation} and~\ref{model:epi} individually and compare their solutions. Generally, it is more efficient to solve these models individually and test for convergence via dynamic callbacks.

\subsection*{Variable initialization}
Defensive Model~\ref{model:epi} can be hard to solve numerically, for both Ipopt (local solution) and Gurobi (finding an incumbent). To overcome this, we initialize, i.e., warm start, the model variables with a locally optimal solution via the following multi-step process. First, we find an injection solution which is naturally ``far" from all feasibility margins by solving
\begin{subequations}
\begin{align}
p_{{\rm init}}=\argmin_{p}\quad & m\\
{\rm s.t.}\quad & a_i^Tp+c_i\le m,\;\forall i\in\mathcal{C}.
\end{align}
\end{subequations}
Next, we initialize the control policy matrix $G$ to be all zeros: $G_{{\rm init}}=0$. Plugging these in to \eqref{eq: epi}, we solve
\begin{align}
t_{{\rm init}}=\max_{t}\quad & t\\
{\rm s.t.}\quad & \left(b_{i}^{T}b_{i}\right)t\le\left(a_{i}^{T}p_{{\rm init}}+c_{i}\right)^{2},\;\forall i\in\mathcal{C},
\end{align}
which is a linear program, to find $t_{{\rm init}}$. %\will{$t$ is the only variable. is this better than writing $t_{\rm init} = \min_{i\in \mathcal{C}} (A_i^T p_{\rm init} + b_i) / (B_i^T B_i)$?} 
We then pass the feasible solution tuple ($p_{{\rm init}}$, $G_{{\rm init}}$, $t_{{\rm init}}$) to Ipopt, which finds a locally optimal nonlinear program (NLP) solution to \eqref{eq: epi}. This locally optimal solution serves as Gurobi's initial incumbent.
The optimal control policy parameters $(p_0, G)$ and maximal defense radius $t$ are determined by Gurobi's solution.

We also warm start Model~\ref{model:minimum_perturbation}. Ipopt tends to have less numerical challenge with this model, but Gurobi can still have a hard time improving the incumbent. To overcome this, we use a short sequence of random Ipopt initializations to try finding better and better incumbent solutions. Looping over five random initializations, we solve each with Ipopt and pass the best solution to Gurobi. Whenever Ipopt faced numerical infeasibility, we lowered its numerical tolerance to 1e-5.

%solving Model~\ref{model:minimum_perturbation} and finding $\delta^*$. The general strategy is this: employ a CPU-based spatial branch and bound solver to solve \eqref{model:minimum_perturbation}. Meanwhile, employ a GPU-based gradient ascent routine, with stochastic restarts, to maximize \eqref{eq: t_tilde}. As $\tilde t$ iteratively grows in size, we pass these RSOC cuts, via \eqref{eq: cut_rsoc} to the spatial branch and bound solver.

%To maximize $\tilde t$ using gradient-based methods on GPUs, we must take the gradient of \eqref{eq: t_tilde}. Since we are differentiating the smallest element in a set, the gradient is technically a subgradinent. In order to compute this subgradient, we first identify the index of the smallest minimal perturbation, and then we take the gradient of the associated minimal perturbation via
%\begin{align}
%\frac{\partial}{\partial\{G,p_{0}\}}\!\!\left(\!\!\tfrac{\left(A_{j}^{T}p_{0}+b_{j}\right)^{2}}{\left(G^{T}A_{j}\!+\!B_{j}\!\right)^{T}\!\left(G^{T}A_{j}\!+\!B_{j}\right)}\!\!\right)\!,\;j\!=\!\argmin_{i\in\mathcal{C}}\;\bigl\Vert\delta_{{\rm lb}}^{(i)}\bigr\Vert_{2}^{2}.
%\end{align}
%Algorithm XX summarize the full solution procedure.

\section{Numerical Test Results}\label{sec:tests}
The primary goal of this section is to demonstrate the applicability of Theorem~\ref{eq: th_simp} to power flow problems of interest. We do so by solving Models~\ref{model:minimum_perturbation} and \ref{model:epi} on PGLib test cases and comparing the resulting bounds. The only direct way to find the smallest adversarial attack $\delta^*$ is by solving Model~\ref{model:minimum_perturbation} to global optimality with a spatial BaB solver. All simulated code is posted publicly on GitHub\footnote{\url{https://github.com/SamChevalier/DCAttack}}.

%In this section, we test the performance of Model~\ref{model:mod4}. Specifically, we test its ability to provide attack size guarantees ...

%The only direct way to find the smallest adversarial attack $\delta^*$ is by solving Model~\ref{model:minimum_perturbation} to global optimality with a spatial BaB solver. In this section, use Gurobi 12 to solve Model~\ref{model:minimum_perturbation} on various PGLib test cases. We then use Gurobi to solve Model~\ref{model:mod4} on the same test cases.

\subsection{Test setup}
We study adversarial load perturbations applied to the \textit{nominal} base load for each test case. At each bus $i$ with nonzero load, we assign a variable load perturbation $\delta_i$. We do not bound or weight these load perturbations in any way\footnote{This assumption can be removed by defining a diagonal weighting matrix, $W$, where $W_{i,i}$ corresponds to inverse load perturbation variances. By updating the object function via $\delta^{T}W\delta$, the minimizer will prioritize scaling loads with large variances in order to reach infeasibility.}. Existing generators can freely change their generation without cost consideration, but they must remain within their bounds, per DC-OPF of Model~\ref{model:dcopf}. Test cases are pulled from the PGLib repository~\cite{Babaeinejadsarookolaee:2019}. All models are solved via Gurobi v12's spatial branch and bound solver~\cite{gurobi} and Ipopt~\cite{wachter2006implementation} (for warm starting) via JuMP~\cite{Lubin2023} in Julia 11.2. We initialize all variables passed to Gurobi's spatial BaB solver via the warm start procedure described in Sec.~\ref{sec:solution_method}

To solve Model~\ref{model:minimum_perturbation} in practicality, we convert the strict inequality constraint \eqref{eq: f2} into an equality constraint, such that
\begin{align}\label{eq: eta_eq}
\mu^{T}\left(B\delta+c\right)=\epsilon,\quad 1\gg\epsilon>0.
\end{align}
The positive parameter $\epsilon$ is chosen large enough to be well above the numerical precision threshold, but small enough as to not affect the solution in any practical way. In our tests, we find that very small values of $\epsilon$ lead to slower branch-and-bound convergence, and large values of $\epsilon$ lead to solution inaccuracies. We use $\epsilon=10^{-3}$, which suffers minimally from either problem.

\subsection{Case study: 5-bus network}
\begin{figure}
\begin{center}
\includegraphics[width=1.0\columnwidth]{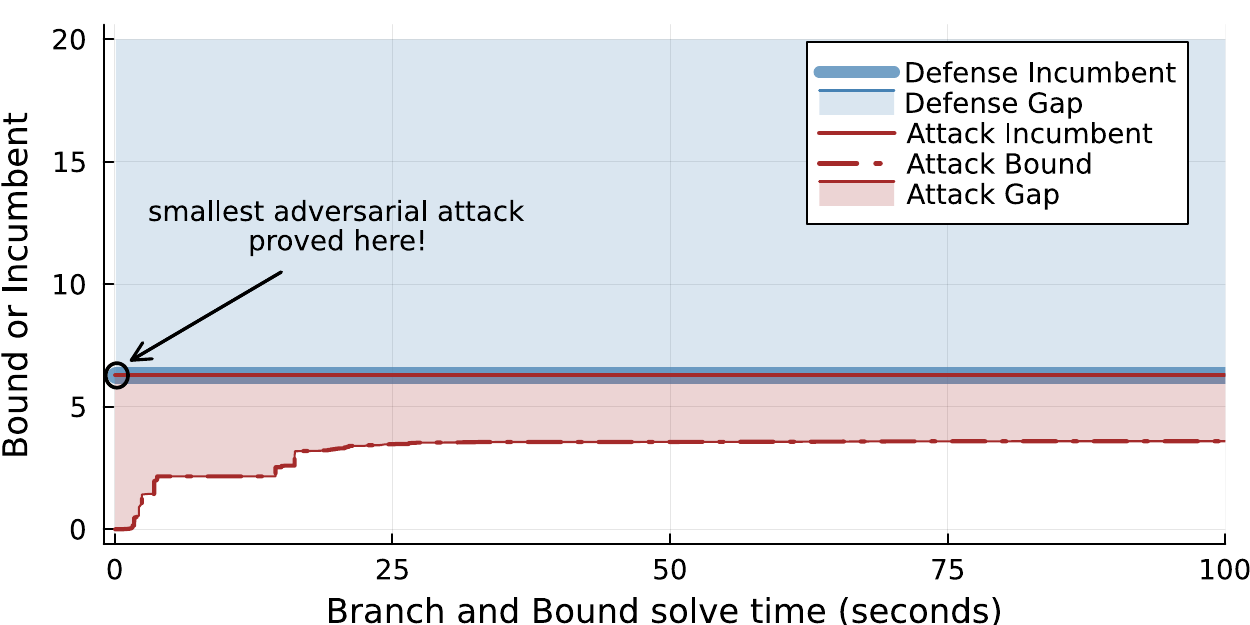}
\end{center}
\caption{Iterative branch and bound solutions (recorded via callbacks) to the 5-bus test case for Models~\ref{model:minimum_perturbation} (attack) and~\ref{model:epi} (defend). Neither model can prove its bound, but the incumbents associated with both models closely match when Gurobi is initialized. %For illustrative purposes, the attack upper bounds are scaled way down from their true values.
}
\label{fig:bab}
\end{figure}

Before presenting test results on the full set of test cases, we apply BaB on the 5-bus test case, solving both Models~\ref{model:minimum_perturbation} and \ref{model:epi}. Using callbacks, we record and plot the incumbent solutions and the upper/lower bounds for both models over time in Fig.~\ref{fig:bab}. As illustrated, at time $T=0$, both incumbents match at a value of 6.29. Since this is the smallest known attack and the largest known defense, this constitutes a global solution. However, unaware of this result, both models continued to BaB, unnecessarily attempting to improve and/or prove their respective upper and lower bounds.

\begin{figure}
\begin{center}
\includegraphics[width=0.95\columnwidth]{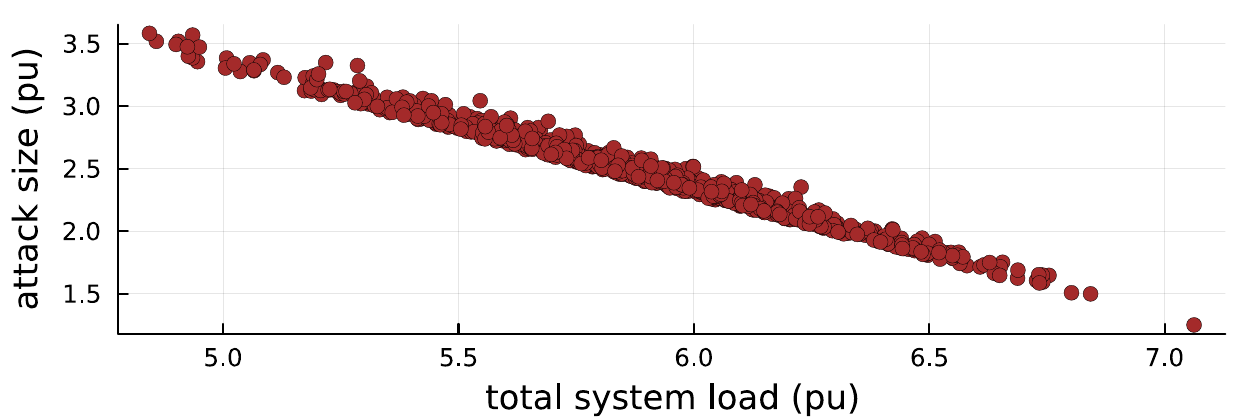}
\end{center}
\caption{Base system load operating sizes $||p_d||_2$ (x-axis) vs the smallest adversarial attack size $||\delta^*||_2$ (y-axis) over 1000 random trials.}
\label{fig:hist}
\end{figure}

To investigate how the smallest adversarial attack size changes as a function of network loading, we solved 1000 adversarial attack problems under 1000 different base loading instantiations. To change the network loads' setpoints, we used a randomly perturbed version of the nominal base load at each bus: $p_{d,i}=p_{d0,i}(1+\nu)$, where $\nu$ is a zero mean Gaussian with a standard deviation of 0.1. The results of this simulation are depicted in Fig.~\ref{fig:hist}. There is an approximately linear relationship between the system load and the attack size (larger system load correlates with smaller adversarial attacks).

%Both distributions are approximately bell-shaped, but the attack distribution has a larger standard deviation than the load distribution. The attack distribution also has an elongated tail on the right, meaning certain load configurations lead to relatively larger defense radii.

\subsection{Larger network test results}
In this subsection, we present results collected on seven PGLib test cases. Specifically, we solved Model~\ref{model:minimum_perturbation} (adversarial attack) and Model~\ref{model:epi} (control defense), running the BsB solver for 30 minutes on each network, for each model (except in the 118 bus case, where the solver ran for 48 hours, as described in Appendix \ref{AppF}). Results are tabulated in Table \ref{tab:compare}. In this table, we present the incumbent and lower bound solutions for Model~\ref{model:minimum_perturbation}, and the incumbent and upper bound solution for Model~\ref{model:epi} (3 significant digits). For the first six test cases, the best incumbent solutions for both models match to within $<1\%$, meaning the globally smallest adversarial attack was successfully identified. The final column shows the time at which these incumbent solutions first matched. Matching incumbents indicates that the defense and the attack have squeezed together, meeting at the uncrossable line in Fig~\ref{fig:Model_bounds}. In all cases but the final one, these incumbents matched within a few seconds. This is due to both ($i$) successful warm starting, and ($ii$) Gurobi being able to improve the incumbent very quickly. No test case achieved a global solution to Model~\ref{model:minimum_perturbation}: most of the lower bounds were stuck right at 0, even after 30 minutes of solve time.

In the largest test case, with 118 buses, the smallest known adversarial attack ($0.449$) was still $\sim10\%$ larger than the largest identified control radius ($0.409$). However, since the lower bound on the adversarial attack was never improved\footnote{To attempt to improve this gap, we ran Gurobi for 12 hours on a computing cluster with 24 CPU cores and 128 GB of memory. Gurobi's lower bound remained at 0.0 for the entire solve time.}, we cannot know from this calculation if there exists a smaller attack vector whose size is between $0.449$ and $0.409$. We do know with certainty, however, that $0.409$ is the largest possible defense radius, since its upper bound was proved by Gurobi. Further details regarding this test and the model reformulations we implemented in an attempt to further close this bound are found in Appendix \ref{AppF}.

\begin{figure}
\begin{center}
\includegraphics[width=0.9\columnwidth]{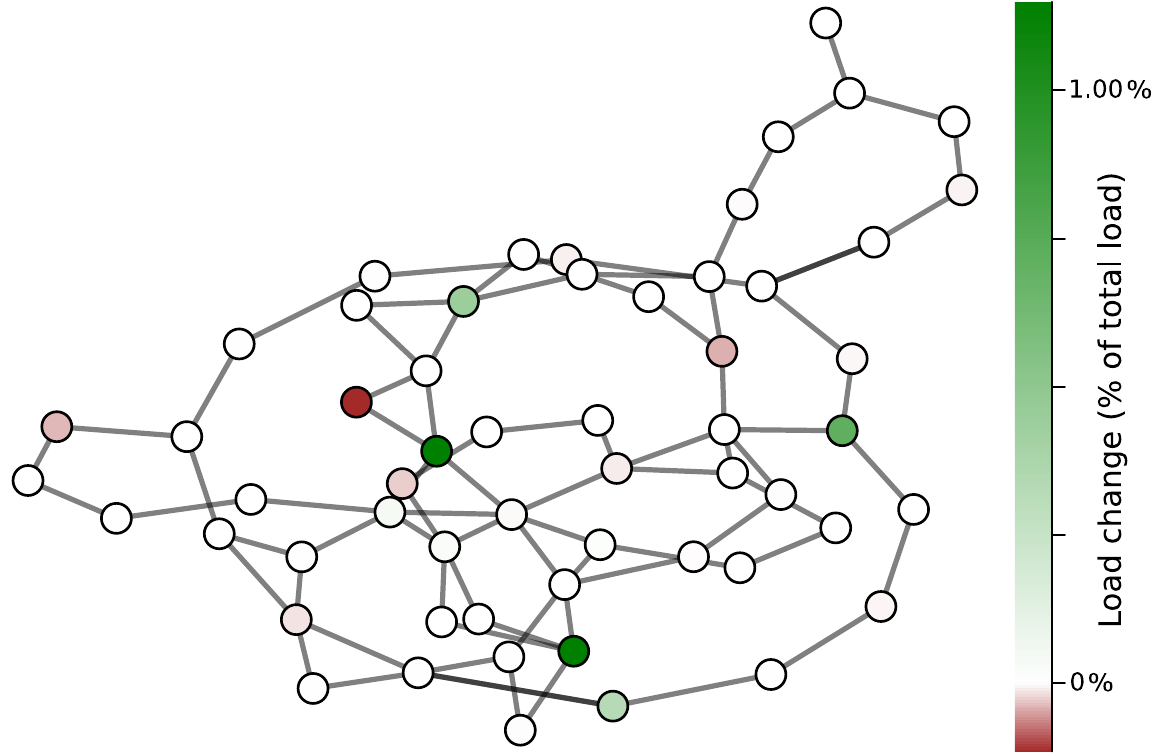}
\end{center}
\caption{Illustration of the smallest adversarial load perturbations needed to drive the 57-bus power grid to the brink of infeasibility. Perturbations are given in percentage of the total system load.}
\label{fig:smallest_attacks}
\end{figure}

We illustrate the smallest adversarial attack identified for the 57-bus network in Fig.~\ref{fig:smallest_attacks}, which was generated using \btt{PowerPlots.jl} \cite{rhodes2025powerplots}. The attack itself is stealthy and surprising, in the sense that a diverse set of loads increase and decrease very slightly to drive the system to infeasibility.

\begin{table}
   \caption{Adversarial Attack and Control Defense Comparisons (30 min BaB)} 
   \label{tab:compare}
   \small
   \centering
   \begin{tabular}{c|ccccc}
   \toprule\toprule
   \multirow{2}{4em}{\textbf{PGLib \\Case}} & \multicolumn{2}{c}{Attack ($\delta^T\delta$)} & \multicolumn{2}{c}{Defend ($t$)} & Match\\ &LB & Incumbent & Incumbent & UB & (sec)\\  

   \midrule 
   $\;$\textit{5\_pjm}   & $5.28$ & \textbf{6.29}  & \textbf{6.29}   & 1e6  & 0.012\\
   $\;$\textit{14\_ieee} & $0.0$ & \textbf{0.178}  & \textbf{0.179}  & 0.29 & 0.018\\
   $\;$\textit{24\_ieee} & $0.0$ & \textbf{1.81}  & \textbf{1.81}   & 1e100 & 0.339  \\
   $\;$\textit{30\_as}   & $0.0$ & \textbf{0.0144}  & \textbf{0.0144}  & 0.0144 & 0.165 \\   $\;$\textit{57\_ieee} & $0.0$ & \textbf{0.0547}  & \textbf{0.0547}   & 10.1  & 0.067\\
   $\;$\textit{60\_c}    & $0.0$ & \textbf{8.87}  & \textbf{8.87}  & 9.64 & 0.33\\
   $\;$\textit{118\_ieee}& $0.0$ & \textbf{0.449}$^{**}$  & \textbf{0.409}$^{**}$  & 0.409 & - \\
\bottomrule
   \end{tabular}
\end{table}

\section{Conclusion}\label{sec:conclusion}

This paper designed methods to identify the smallest load perturbation which renders DC-OPF constraints infeasible. This problem, which is inherently nonconvex, was formulated using an adversarial attack framework. A parameterized version of Farkas' lemma was used to model this problem, but proving the lower bound tended to be very hard for Gurobi. To overcome this, we proposed an optimizable control policy framework which ``defends" against adversarial attacks, thus providing solvability guarantees. Tests run on small PGLib test cases showed promising early results. Incumbent solutions could reliably find, and prove, the globally smallest adversarial attack in most cases. In the 118-bus system, 30 minutes of branching and bounding was unable to prove the smallest adversarial attack size; however, our method provided a lower bound corresponding to $\sim$10\% optimality gap.

While initial results are promising, the proposed methods will not scale well to networks with many thousands of nodes. This is for three primary reasons. First, the proposed formulation exploits a dense PTDF matrix in the inequality constraints, thus losing all network sparsity and associated efficiency; second, the affine control matrix $G$ is generally dense, so the number of decision variables becomes intractable in higher dimensions; and third, the proposed models are nonconvex, so finding good, ``stealthy" incumbents that close the optimality gap will get harder and harder in higher dimensional search spaces. While scalability is the primary challenge associated with the methods presented in this paper, we believe our theoretical contributions will help lay the groundwork for future work that can quickly and efficiently find stealthy adversarial attacks in realistically sized power grids.

Future work will overcome these limitations to boost the scalability of the proposed approach and extend these results to the nonlinear AC-OPF context. This will require adapting our methods to directly handle equality constraints (i.e., rather than solving them away), but it could open the door to many interesting industry-relevant applications, as explained in the following subsection. Finally, we note that the proposed control policy $p=p_0+G\delta$ was the best affine control policy we could identify. However, there exist an infinite number of alternative control policies (e.g., piecewise linear control policies, quadratic control policies, etc.) which could broaden the applicability of our results (e.g., to the nonlinear AC-OPF context). Future work will explore these alternative control policies.

\subsection*{Potential industry applications}

Identifying the smallest perturbation that engenders constraint infeasibility has many potential industry applications.
\begin{itemize}
    \item To ensure a robust day-ahead market clearing solution, system operators may want to ensure their DC-OPF solutions are at least $\epsilon$-robust to load forecast uncertainty.
    \item Similarly, given renewable energy dispatch uncertainty, system operators may want to ensure their generator unit commitment decisions are sufficiently robust.
    \item Finally, if a system operator is going to switch network lines, they might want to co-optimize both the cost of network operation with the size of the network robustness margin. Line switching decisions that result in low operating costs, but a very small margin to network infeasibility, may be considered disadvantageous. 
\end{itemize}

Model solve time requirements will depend on the intended application (e.g., order of minutes for real-time market clearing, v.s., order of hours or days for unit commitment).

\section{Acknowledgments}\label{sec:ack}
The authors thank Hassan Hijazi of Gurobi who provided helpful reformulation recommendations for Model~\ref{model:minimum_perturbation}.

%recommendation regarding the reformulation of Model~\ref{model:minimum_perturbation} to improve BaB solve time.

\appendices
\renewcommand{\baselinestretch}{.9}
{\section{}\label{AppA}}
Without loss of generality, we assign generator 1 as the slack generator. Solving the power balance equation via
\begin{align}
p_{g,1}=1^{T}\left(p_{d}+\delta\right)-\hat{1}^{T}\hat{p}_{g},
\end{align}
the generation limit and flow limit inequalities are given as
\begin{align}
\underline{p}_{f}\leq\hat{\Phi}\hat{p}_{g}-\Phi\left(p_{d}+\delta\right) & \leq\overline{p}_{f}\\
\underline{p}_{g}\leq\left[\begin{array}{c}
1^{T}\left(p_{d}+\delta\right)-\hat{1}^{T}\hat{p}_{g}\\
\hat{p}_{g}
\end{array}\right] & \leq\overline{p}_{g}.
\end{align}
Thus, the following compact inequality can be formulated:
\begin{align}
\underbrace{\left[\!\!\begin{array}{c}
\hat{\Phi}\\
-\hat{\Phi}\\
-\hat{1}^{T}\\
I\\
\hat{1}^{T}\\
-I
\end{array}\!\!\right]}_{A}p+\underbrace{\left[\begin{array}{c}
-\Phi\\
\Phi\\
1^{T}\\
0\\
-1^{T}\\
0
\end{array}\right]}_{B}\delta+\underbrace{\left[\!\!\begin{array}{c}
-\Phi p_{d}-\overline{p}_{f}\\
\underline{p}_{f}+\Phi p_{d}\\
\left[\!\!\begin{array}{c}
1^{T}p_{d}\\
0
\end{array}\!\!\right]-\overline{p}_{g}\\
\underline{p}_{g}-\left[\!\!\begin{array}{c}
1^{T}p_{d}\\
0
\end{array}\!\!\right]
\end{array}\!\!\right]}_{c}\le0.
\end{align}
To simplify paper notation, we have defined $p\triangleq \hat{p}_g$ as the reduced generation vector. We note that $\hat{\Phi}\hat{p}_{g}$ and ${\Phi}{p}_{g}$ are equivalent, since slack generator injections do not explicitly alter network flows. Finally, if load perturbations or generators are only present at subsets of buses, we can introduce matrices, $N_g$ and $N_d$, which map load and generation to their respective buses via, e.g., $\Phi N_gp_{g}$ and $\Phi N_dp_{d}$. Such mapping matrices are used in the numerical test results sections.

{\section{}\label{AppB}}
Taking the Lagrange dual of \eqref{eq: quad_proj} yields
\begin{align}
\max_{\lambda}\min_{\delta}\;\delta^{T}\delta+\lambda\left(a_{i}^{T}p_0+b_{i}^{T}\delta+c_{i}\right).
\end{align}
Applying stationarity conditions to the primal, via $\frac{\partial\mathcal{L}}{\partial\delta}=2\delta+\lambda b_{i}=0$ yields a solution $\delta=-\frac{1}{2}\lambda b_{i}$. Plugging this back into the Lagrangian yields
\begin{align}
\max_{\lambda}\;-\frac{1}{4}\lambda^{2}b_{i}^{T}b_{i}+\lambda\left(a_{i}^{T}p_0+c_{i}\right).
\end{align}
Again applying stationarity conditions, via $\frac{\partial\mathcal{L}}{\partial\lambda}=-\frac{1}{2}\lambda b_{i}^{T}b_{i}+a_{i}^{T}p_0+c_{i}=0$, yields a solution for $\lambda$:
\begin{align}
\lambda=2\frac{a_{i}^{T}p_0+c_{i}}{b_{i}^{T}b_{i}}.
\end{align}
Plugging this solution into $\delta=-\frac{1}{2}\lambda b_{i}$ yields
\begin{align}
\delta^{(i)}_{\rm lb}=-\frac{\left(a_{i}^{T}p_0+c_{i}\right)}{b_{i}^{T}b_{i}}b_{i}.
\end{align}
%\will{Was it obvious a priori that $\delta_{\rm lb}^{i}$ is a projection of $a_i^Tp_0 + c_i$ onto $b_i$? If it's obvious, maybe we should mention it. If not, ... is it interesting?}
Finally, we may take the inner product of $\delta^{(i)}_{\rm lb}$ with itself to compute the size of the perturbation: 
\begin{align}\label{eq: pi_solution}
(\delta^{(i)}_{\rm lb})^T\delta^{(i)}_{\rm lb}=\frac{\left(a_{i}^{T}p_0+c_{i}\right)^{2}}{b_{i}^{T}b_{i}}.
\end{align}
We may directly extend this solution to account for inclusion of the parameterized control policy $G$ in \eqref{eq: p_control}:
\begin{align}\label{eq: ld_gp}
\max_{\lambda}\min_{\delta}\;\delta^{T}\delta+\lambda\left(a_{i}^{T}p_{0}+\left(a_{i}^{T}G+b_{i}^{T}\right)\delta+c_{i}\right).
\end{align}
Its solution is given by substituting into \eqref{eq: pi_solution},
\begin{align}
\delta^{(i)}_{\rm lb}=-\frac{\left(a_{i}^{T}p_0+c_{i}\right)}{\left(G^{T}a_{i}+b_{i}\right)^{T}\left(G^{T}a_{i}+b_{i}\right)}\left(G^{T}a_{i}+b_{i}\right).
\end{align}
We may take the inner product of $\delta^{(i)}_{\rm lb}$ with itself to compute the size of this perturbation: 
\begin{align}\label{eq: param_pert_size}
\bigl\Vert \delta_{{\rm lb}}^{(i)}\bigr\Vert _{2}^{2}=\frac{\left(a_{i}^{T}p_0+c_{i}\right)^{2}}{\left(G^{T}a_{i}+b_{i}\right)^{T}\left(G^{T}a_{i}+b_{i}\right)}.
\end{align}

{\section{}\label{AppC}}
For a general linear program with feasibility constraints
\begin{subequations}\label{eq: f1a}
\begin{align}
Ax+b & =0\\
Cx+d & \le0,
\end{align}
\end{subequations}
Farkas' lemma offers an alternative system of equations
\begin{subequations}\label{eq: f2a}
\begin{align}
\lambda^{T}A+\mu^{T}C & =0\\
\lambda^{T}b+\mu^{T}d & >0\\
\mu  &\ge 0.
\end{align}
\end{subequations}
Either \eqref{eq: f1a} or \eqref{eq: f2a} is satisfiable, but never both.

{\section{}\label{AppD}}
\textbf{Example 1:} \textit{Rank-1 Control Policy. }Assuming a uniform distributed slack model, injection perturbations ($p-p_0$) are mapped to nodal generator responses via 
\begin{align}
\Phi\left(p+1\frac{1^{T}(p-p_{0})}{n}\right)=\underbrace{\Phi\left(I+\frac{11^{T}}{n}\right)}_{\text{rank-1 PTDF update}}p-\underbrace{\Phi\frac{11^{T}}{n}p_{0}}_{\rm bias}.
\end{align}
Thus, distributed slack implicitly applies a rank-1 update to the PTDF matrix. \hfill $\qed$

{\section{}\label{AppE}}
\begin{model}[h]
\caption{\hspace{-0.1cm}\textbf{:} Squeezing the Attack and Defense}
\label{model:mod4}
\vspace{-0.25cm}
\begin{align*}
\tau^{*}\triangleq\min_{\mu\ge0,\delta,t,G,p_{0}}\quad & \delta^{T}\delta - t\\
{\rm s.t.}\quad&\,\eqref{eq: f1}\text{ - }\eqref{eq: f2}\tag{Farkas' lemma}\\
&\,\eqref{eq: tle}\text{ - }\eqref{eq: local_const}\tag{Control policy}
\end{align*}
\vspace{-0.5cm}
\end{model}

At optimality, if $\tau^{*}=0$, an affine control policy has been found that generates feasible primal solutions $\forall \delta \in \Delta$. If $\tau^{*}\ne0$, then no such policy exists, but $\tau^{*}$ provides a gap between the smallest adversarial attack and the best affine control policy. %Either way, at optimality, the value of $\delta^T\delta$ will be the smallest possible adversarial attack, and $t$ will be the best possible defense radius. 

{\section{}\label{AppF}}
Closing the gap between the optimal defense radius and the smallest identified adversarial attack
on the 118 bus test case was challenging. In attempt to decrease the gap, we tested three different adversarial attack models: ($i$) Model~\ref{model:minimum_perturbation}, based on Farka's lemma, ($ii$) Model \ref{model:minimum_perturbation} with the added constraint $\sum_i\mu_i = 1$, and ($iii$) an alternative framework, which directly defines an infeasible perturbation via the bilevel optimization problem
\begin{subequations}
\begin{align}
\min_{\delta,t}\quad&\delta^{T}\delta\\
{\rm s.t.} \quad&t>0\\
 & t=\min_{p,\epsilon}\;\;\epsilon\label{eq: epsilon}\\
 & \qquad\;{\rm s.t.}\;\; Ap+B\delta+c\le\bm{1}\epsilon,
\end{align}
\end{subequations}
where $\bm{1}$ is the vector of all ones. 
The inner loop tries to avoid violations by minimizing (over $p$) the largest constrained quantity of \eqref{eq: feas_space}, while the outer loop forces the lower-level problem to violate a constraint by requiring $\epsilon > 0$. 
Reformulating the lower level with the Karush-Kuhn-Tucker conditions~\cite{boyd2004convex}, we arrive at the single-level adversarial attack problem
\begin{subequations}\label{eq: reformulation}
\begin{align}\min_{\delta,\epsilon,p,\mu\ge0}\quad & \delta^{T}\delta\\
{\rm s.t.}\quad & \epsilon>0\\
 & Ap+B\delta+c\le\bm{1}\epsilon\\
 & \mu_{i}\left(Ap+B\delta+c-{\bm 1}\epsilon\right)_{i}=0,\forall i\label{eq: compslack}\\
 & A^{T}\mu=0\label{eq: stat1}\\
 & \mu^{T}\bm{1}=1\label{eq: stat2},
\end{align}
\end{subequations}
where $t$ has been replaced by $\epsilon$ per \eqref{eq: epsilon}.
Line \eqref{eq: compslack} is complementary slackness, while \eqref{eq: stat1}-\eqref{eq: stat2} are stationarity conditions (with respect to $p$ and $\eta$, respectively). 
We note that Model~\ref{model:minimum_perturbation} can be expressed as a relaxation of \eqref{eq: reformulation}: we keep the constraint \eqref{eq: stat1} and sum the constraints \eqref{eq: compslack} over $i$ to get $\mu^T(B\delta + c) = \epsilon$, which is identical to the constraint in Model~\ref{model:minimum_perturbation}.

In our implementation, we tested $\epsilon=10^{-3}$, $10^{-4}$, and $10^{-5}$ for numerical sensitivity; we observed that achieving accurate solutions with the addition of \eqref{eq: stat2}, which essentially normalizes $\mu$, required smaller values of $\epsilon$ relative to the value used in Model \ref{model:minimum_perturbation}. All three attack formulations ran for 48 hours on 24 HPC CPU cores with 128GB of memory. Table \ref{tab:compare_attacks} lists the smallest attacks found with each model.

%While \eqref{eq: reformulation} is still a hard problem, Gurobi's branch and bound solver was able to find better incumbent more quickly than it could for Model \ref{model:minimum_perturbation}. The following table summarizes our results

\begin{table}
   \caption{Best Adversarial Attack Solutions (48hr BaB)} 
   \label{tab:compare_attacks}
   \small
   \centering
   \begin{tabular}{c|ccc}
   \toprule\toprule
   118-bus & Mod \ref{model:minimum_perturbation} & Mod \ref{model:minimum_perturbation} + \eqref{eq: stat2} & Bilevel \eqref{eq: reformulation}\\  

   \midrule 
   $\;$Smallest Attack   & $0.580$ & $0.449$  & $0.449$ \\
   $\;$ Bound Reached (hr)   & 0.01 & 12.97  & 2.64\\
\bottomrule
   \end{tabular}
\end{table}

\bibliographystyle{IEEEtran}
\bibliography{references}

\end{document}